\documentclass[
showkeys,	
reprint,	
amsmath,	
amssymb,	
amsfonts,	
aps,		
pre,%
groupedaddress,
superscriptaddress,
a4paper,%
nofootinbib,
eprint,		
final,		
preprintnumbers, 
]{revtex4-2}

\usepackage[tbtags]{mathtools}
\usepackage{amsmath,amsfonts,mathdots,amssymb,yfonts,calc}
\usepackage{graphicx,graphics,xcolor}
\usepackage{subfigure}

\usepackage{natbib}
\usepackage{booktabs}
\usepackage{siunitx}
\usepackage{float}
\usepackage{tabularray}
\usepackage{multirow}
\usepackage{physics}
\usepackage{dcolumn}
\usepackage[utf8]{inputenc}  
\usepackage[T1]{fontenc} 
\usepackage{natbib}
\usepackage{booktabs}
\usepackage{hyperref}
\hypersetup{
    colorlinks=true,
    linkcolor=blue,
    filecolor=magenta,      
    urlcolor=blue!80,
    citecolor=cyan,
    linktocpage=true,
    breaklinks=false,
}

\newcommand{\ex}{\mathrm{ex}}
\newcommand{\Prob}{\mathbb{P}}
\newcommand{\Ffun}{\mathcal{F}}
\newcommand{\KD}{\mathcal{K}}
\newcommand{\GD}{\mathcal{G}}
\newcommand{\LR}{\mathrm{LR}}
\newcommand{\ssstate}{\mathrm{ss}}
\newcommand{\Lag}{\mathcal{L}}
\begin{document}
\title{Long-Range Nonequilibrium Correlations as a Thermodynamic Speedometer}
\author{Vinesh Vijayan}
\email{vinesh.physics@rathinam.in}
\affiliation{Department of Science and Humanities, Rathinam Technical Campus, Coimbatore, India -641021}
\author{Rishma Thilakaraj}
\affiliation{Department of Science and Humanities, Rathinam Technical Campus, Coimbatore, India -641021}
\author{Gokila V}
\affiliation{Department of Science and Humanities, Rathinam Technical Campus, Coimbatore, India -641021}
\author{ Yogeshvari V A}
\affiliation{Department of Science and Humanities, Rathinam Technical Campus, Coimbatore, India -641021}

\date{\today}

\begin{abstract}
Despite purely local microscopic dynamics, nonequilibrium steady states can develop long-range correlations. We investigate this phenomenon in the boundary-driven symmetric simple exclusion process by relating its exact nonlocal large-deviation functional to finite-time thermodynamic irreversibility under weak time-dependent driving. Decomposing the long-range correlations into hydrodynamic modes, we find that the lag associated with the slowest mode satisfies an upper bound determined by the excess entropy production rate at each driving frequency. This bound is approached in the quasistatic limit, whereas progressively shorter-wavelength modes do not generally satisfy the corresponding bound. These results provide a quantitative connection between the curvature of the large-deviation functional, long-range correlations in the nonequilibrium steady state, and finite-time dissipation. Within the linear-response regime considered here, they indicate that the longest-wavelength hydrodynamic mode has a distinct relation between slow relaxation and thermodynamic irreversibility.

\end{abstract}

\maketitle
\pagestyle{plain}

\section{\label{s1}Introduction}
Even when the underlying microscopic dynamics are strictly local, nonequilibrium steady states (NESSs) in driven diffusive systems can exhibit spatially long-ranged correlations. Such correlations are a characteristic signature of nonequilibrium driving and have been extensively studied within the framework of fluctuating hydrodynamics and macroscopic fluctuation theory \cite{Spohn1983,Bertini2015,Bertini2002}. An exact calculation demonstrated the same phenomenon for the boundary-driven symmetric simple exclusion process (SSEP). In particular, Derrida, Lebowitz, and Speer showed that the probability of observing a density profile $\rho(x)$ in a system of size $N$ scales as
$e^{-N\mathcal{F}[\rho]}$, where $\mathcal{F}[\rho]$ is a nonlocal nonequilibrium free-energy functional \cite{Derrida2001,Derrida2002,Bertini2003}. The long-range correlations of the NESS are directly reflected in the nonlocal structure of this functional. Macroscopic fluctuation theory (MFT) provides a unified hydrodynamic framework for describing density and current fluctuations in driven diffusive systems, within which this large-deviation structure naturally arises \cite{Bertini2015,Bertini2002}.

A natural dynamical question is whether the nonlocal curvature of the nonequilibrium free-energy functional constrains the time-dependent response of correlations when the system is driven at a finite rate. By treating the correlation field itself as a fluctuating thermodynamic observable, Bodineau and Dagallier recently established a large-deviation framework for time-averaged two-point correlations in the open SSEP \cite{Bodineau2024}. In parallel, nonequilibrium thermodynamics has developed a systematic decomposition of entropy production into excess and housekeeping components \cite{Hatano2001,Maes2014}. More recent generalized free-energy constructions have connected large deviations, excess/housekeeping dissipation, and thermodynamic speed limits \cite{Kolchinsky2026}. Thermodynamic bounds have also been established for the intrinsic correlation times of observables and for the speed of stochastic processes in terms of entropy production \cite{Dechant2023,Shiraishi2018}. These developments point toward a possible direct connection between the lag of long-range correlations and the irreversible cost of driving. Yet, a quantitative relation between the finite-frequency lag of individual correlation modes and excess entropy production has not yet been established.

Using fluctuating hydrodynamics to connect the Derrida--Lebowitz--Speer (DLS) large-deviation functional to the finite-frequency response, we address this question for the boundary-driven symmetric simple exclusion process (SSEP). The boundary density difference is weakly modulated harmonically as $g(t)=g_0+\epsilon\cos(\omega t)$, and the correlation dynamics are resolved in the Dirichlet eigenmodes of the diffusive operator. The stationary DLS functional yields the correlation-mode spectrum $a_n^{\mathrm{ss}}=g_0^2/(n\pi)^2$, with the corresponding long-range covariance kernel proportional to the inverse Laplacian. Under time-dependent driving, the correlation modes acquire a frequency-dependent lag $\ell_n(\omega)$, with the low-frequency behavior scaling as $\ell_n(\omega)\sim\omega/(n\pi)^4$.

We find that the softest mode, $n=1$, satisfies the proposed dissipation bound over the full frequency range considered, whereas higher spatial modes systematically violate the corresponding inequality. The bound involves only the excess component of dissipation—the irreversible cost associated with driving the system away from its instantaneous stationary state—consistent with the excess/housekeeping framework \cite{Hatano2001,Maes2014,Kolchinsky2026}. The result identifies the longest-wavelength correlation mode as a thermodynamically distinguished speedometer. It establishes a direct link between the curvature of a nonequilibrium free-energy functional, the dynamics of long-range correlations, and excess dissipation, thereby connecting large-deviation geometry to finite-time irreversibility.

The paper is organized as follows. In Sec. \ref{s2}, we introduce the boundary-driven symmetric simple exclusion process (SSEP) and the weak time-dependent boundary driving, followed by the large-deviation functional and the associated long-range correlation kernel. We then resolve the correlation field into the Dirichlet eigenmodes of the diffusive operator and obtain the finite-frequency response and mode-dependent correlation lag. In Sec. \ref{s3}, we introduce the excess entropy production associated with the time-dependent driving and derive the correlation speed limit. We then examine the frequency and mode dependence of the bound, with particular emphasis on the softest mode and the high-frequency violations observed for higher spatial modes. In Sec. \ref{s4}, we discuss the main results, their physical interpretation, their relation to the broader nonequilibrium framework, and the limitations and possible extensions of the present analysis. Detailed derivations of the large-deviation functional, correlation kernel, mode dynamics, finite-frequency response, excess entropy production, speed-limit inequality, and high-frequency behavior are provided in the Appendices.
\section{\label{s2}Model and DLS correlation kernel}

Consider a diffusive SSEP on the interval $(0<x<1)$, with coarse-grained density $(\rho(x,t))$. We adopt the convention
\begin{equation}
D=1,\qquad \sigma(\rho)=2\rho(1-\rho),
\label{E2}
\end{equation}
where the factor of 2 fixes our mobility convention. We denote by $(\rho_0)$ the mean density and by $(g(t))$ the density difference between the two reservoirs. The boundary densities are defined as
\begin{equation}
\rho_L(t)=\rho_0-\frac{g(t)}{2},
\qquad
\rho_R(t)=\rho_0+\frac{g(t)}{2},
\label{E3}
\end{equation}
so that the mean density remains fixed at $(\rho_0)$, while only the density difference between the two boundaries is varied. We consider the weakly driven regime, $(|g|\ll1)$, sufficiently close to equilibrium for linear-response perturbation theory to apply. The equilibrium compressibility of the SSEP is
\begin{equation}
\chi_0=\rho_0(1-\rho_0),
\label{E4}
\end{equation}
which is also the local variance of the occupation variable, since each SSEP site is effectively a Bernoulli variable.

According to the Derrida–Lebowitz–Speer (DLS) large-deviation result, for fixed boundary densities, the probability of observing a density profile $(\rho(x))$ in a system of size $(N\gg1)$ is given by
\begin{equation}
\Prob[\rho]\asymp \exp[-N\Ffun[\rho]],
\label{E5}
\end{equation}
where
\begin{equation}
\Ffun[\rho]
=
\sup_F
\int_0^1\dd x
\left[
\rho\ln\frac{\rho}{F}
+(1-\rho)\ln\frac{1-\rho}{1-F}
+\ln\frac{F'}{g}
\right],
\label{E6}
\end{equation}
with the supremum taken over an auxiliary field (F(x)) satisfying the boundary conditions
\begin{equation}
F(0)=\rho_L,\qquad
F(1)=\rho_R.
\label{E7}
\end{equation}
Derivation of the functional and its first variation is given in Appendix \ref{A1}. The maximizing field satisfies
\begin{equation}
\rho
=
F+F(1-F)\frac{F''}{(F')^2}.
\label{E8}
\end{equation}

For the SSEP, the typical stationary density profile is linear, consistent with the vanishing of its second derivative. In the present setting, it is given by
\begin{equation}
\bar\rho(x)=\rho_0+g\left(x-\frac12\right),
\label{E9}
\end{equation}
for which the maximizing field is simply $(F=\bar\rho)$.
The functional derivative of $\mathcal{F}$ with respect to $\rho$ is obtained using the envelope theorem:
\begin{equation}
\frac{\delta \mathcal{F}}{\delta \rho(x)}
=
\ln\frac{\rho(x)[1-F(x)]}{F(x)[1-\rho(x)]}.
\label{E10}
\end{equation}
The first variation therefore vanishes at $\rho=\bar{\rho}$.

The next step is to study the fluctuations around the typical profile as follows
\begin{equation}
\rho=\bar\rho+\eta,\qquad
F=\bar\rho+u.
\label{E11}
\end{equation}
where $\eta(x)$ is the flactuations of the physical density while $u(t)$ is the corresponding flactuations of the auxilary field. The goal is the derivation of the Hessian of the large deviation functional. Linearization of Eq.~\eqref{E6} gives
\begin{equation}
\eta = u+\frac{\bar\rho(1-\bar\rho)}{g^2}u'',
\label{E12}
\end{equation}
Both $\rho$ and $F$ have the same fixed boundary values and therefore flactuations vanishes at the boundaries with $u(0)=u(1)=0$.  Let $G_F$ be the Green function of the corresponding linear operator.  Then
\begin{equation}
\frac{\delta^2\Ffun}{\delta\rho(x)\delta\rho(y)}
=
\frac{\delta(x-y)}
{\bar\rho(x)[1-\bar\rho(x)]} -G_F(x,y).
\label{E13}
\end{equation}
the Hessian consists of a long term minus a non local green function. Since the probability is proportional to $e^{-N\mathcal{F}}$, the curvature is proportional to the inverse of the Hessian. The inverse Hessian is
\begin{equation}
\KD^{-1}(x,y)
= \bar\rho(x)[1-\bar\rho(x)]\delta(x-y) -g^2\GD(x,y),
\label{E14}
\end{equation}
where
\begin{equation}
-\partial_x^2\GD(x,y)=\delta(x-y),
\qquad
\GD(0,y)=\GD(1,y)=0,
\label{E15}
\end{equation}
and explicitly
\begin{equation}
\GD(x,y)=
\min(x,y)\,[1-\max(x,y)].
\label{E16}
\end{equation}
Eq.~\eqref{E14} displays the local equilibrium susceptibility and the nonlocal NESS contribution. The Hessian calculation is detailed in Appendix \ref{B1}.

\subsection{Correlation modes and Finite-frequency correlation response}

We introduce the normalized Dirichlet eigenfunctions of the Laplacian,
\begin{equation}
\phi_n(x)=\sqrt{2}\sin(n\pi x),
\qquad
\lambda_n=(n\pi)^2,
\label{E17}
\end{equation}
which form the natural eigenbasis for the problem.
We define the positive long-range correlation amplitude
\begin{equation}
a_n(t)
=
-\int_0^1 \dd x\int_0^1 \dd y,
\phi_n(x)C_{\LR}(x,y,t)\phi_n(y).
\label{E18}
\end{equation}
The quantity $(a_n(t))$ measures the magnitude of the long-range anticorrelation projected onto spatial mode $(n)$. Since
\begin{equation}
G_D(x,y)
= \sum_{n=1}^{\infty} \frac{\phi_n(x)\phi_n(y)}{\lambda_n},
\label{E19}
\end{equation}
the stationary DLS result is [see Appendix \ref{C1}]
\begin{equation}
a_n^{\mathrm{ss}}(g)= \frac{g^2}{\lambda_n}.
\label{E20}
\end{equation}
The factor $(1/\lambda_n)$ arises from the inverse-Laplacian form of the Green's function. 
This immediately yields the algebraic correlation spectrum
\begin{equation}
a_n^{\mathrm{ss}}\sim n^{-2}.
\label{E21}
\end{equation}
Thus, the correlation amplitude decays algebraically with the spatial mode number. The $(n=1)$ mode corresponds to the longest-wavelength spatial mode, while higher modes are progressively suppressed as $(n^{-2})$.
\begin{figure}[H]
    \centering
    \includegraphics[width=0.9\columnwidth]{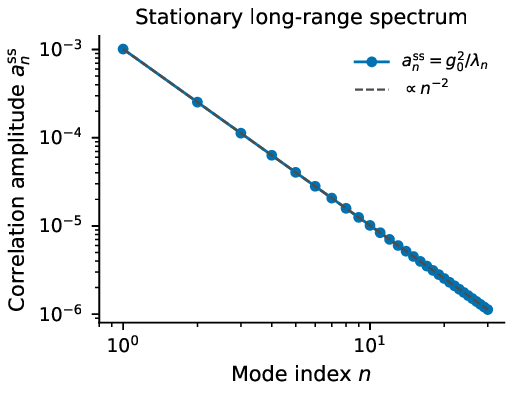}
    \caption{This algebraic spectrum originated from the nonlocal curvature of the DLS large-deviation functional and is the static starting point for the dynamical analysis that follows.}
    \label{fig1}
\end{figure}

Fig.~\ref{fig1} demonstrates that, for the boundary-driven SSEP, the long-range correlation spectrum exhibits a decay of the mode amplitudes (blue dots) as $n^{-2}$, in agreement with the dashed lines. The dominant correlation weight is carried by the longest-wavelength mode, while higher modes are progressively suppressed.

How do correlations respond to periodically varying boundary driving? In particular, do they follow the drive instantaneously, or do they exhibit a finite dynamical lag? To address this question, we modulate the boundary drive as
\begin{equation}
g(t)=g_0+\epsilon\cos(\omega t),
\qquad
\epsilon\ll g_0\ll1.
\label{E22}
\end{equation}
Starting from a weakly driven NESS characterized by the mean gradient $(g_0)$, the modulation periodically increases and decreases the boundary drive. The parameter $(\epsilon)$ sets the small modulation amplitude, while $(\omega)$ is the driving frequency. Since $(\epsilon\ll g_0)$, we retain only terms linear in $(\epsilon)$.

When the boundary drive oscillates, the density in the bulk acquires a time-dependent component. To linear order in the modulation amplitude, the deterministic density takes the harmonic form
\begin{equation}
\rho(x,t)
=
\bar\rho(x)+\epsilon\Re\left[U_\omega(x)e^{i\omega t}\right],
\label{E23}
\end{equation}
where $(U_\omega(x))$ is the complex amplitude of the density response. Its magnitude sets the response amplitude, while its phase encodes the lag relative to the driving. The linearized diffusion equation gives
\begin{equation}
U_\omega''=i\omega U_\omega,
\qquad
U_\omega(0)=-\frac{1}{2},
\quad
U_\omega(1)=\frac{1}{2}.
\label{E24}
\end{equation}

Thus
\begin{equation}
U_\omega(x)
=
\frac{\sinh[k(x-\frac12)]}{2\sinh(k/2)},
\qquad
k=\sqrt{i\omega}.
\label{E25}
\end{equation}
The response is not simply a spatially varying sinusoid; it has both an amplitude and a phase. At low driving frequency, the system has sufficient time to adjust to the changing boundary conditions, and the response is therefore nearly instantaneous. At high frequency, diffusion cannot keep pace with the rapidly varying drive, resulting in an increasingly delayed response. The same competition between the driving frequency $(\omega)$ and the relaxation rate $(\lambda_n)$ governs the correlation response.
The deterministic density describes the evolution of the mean profile, but the fluctuating hydrodynamic description is required to characterize the fluctuations around it. The covariance $(C(x,y,t))$ obeys [a derivation of this covariance equation from the fluctuating hydrodynamic field theory is given in Appendix \ref{D1}].
\begin{equation}
\partial_t C
=(\partial_x^2+\partial_y^2)C
+
2\partial_x\partial_y
\left[\chi(\rho)\delta(x-y)\right].
\label{E26}
\end{equation}

Subtracting the local equilibrium susceptibility and projecting the covariance equation onto the Dirichlet eigenmodes yields the long-range correlation response [see Appendix \ref{E1}]
\begin{equation}
\delta a_n(\omega)
= \frac{
16g_0\epsilon\lambda_n
}{
(2\lambda_n+i\omega)(4\lambda_n+i\omega)
}.
\label{E27}
\end{equation}
This expression quantifies the response of the $(n)^{th}$ correlation mode to a periodic modulation of the boundary drive. In the quasistatic limit, $(\omega\to0)$, the system follows the boundary drive adiabatically, yielding the instantaneous NESS response
\begin{equation}
\delta a_n^{\mathrm{ss}}
=\frac{2g_0\epsilon}{\lambda_n}.
\label{E28}
\end{equation}
\subsection{Correlation lag}
At finite frequency, the correlation response generally differs from the instantaneous NESS response. We therefore define the correlation lag as [see Appendix \ref{E1}]
\begin{equation}
\ell_n(\omega)
=\delta a_n(\omega)-\delta a_n^{\mathrm{ss}}.
\label{E29}
\end{equation}
Explicitly,
\begin{equation}
\ell_n(\omega)
=-\frac{
2g_0\epsilon\omega(\omega-6i\lambda_n)
}{
\lambda_n(\omega-2i\lambda_n)(\omega-4i\lambda_n)
},
\label{E30}
\end{equation}
which quantifies the deviation of the correlation response from the instantaneous NESS value and hence the failure of the correlations to track the time-dependent drive adiabatically and hence
\begin{equation}
|\ell_n(\omega)|^2
=
\frac{
4g_0^2\epsilon^2\omega^2(\omega^2+36\lambda_n^2)
}{
\lambda_n^2(\omega^2+4\lambda_n^2)(\omega^2+16\lambda_n^2)
}.
\label{E31}
\end{equation}
In the quasistatic limit, $(\omega\ll\lambda_n)$, the correlation lag has the expansion
\begin{equation}
\ell_n(\omega)
= -\frac{3i g_0\epsilon\omega}{2\lambda_n^2}
+O(\omega^2).
\label{E32}
\end{equation}
Thus, the leading finite-frequency lag scales as $(\lambda_n^{-2}\sim n^{-4})$, indicating that higher spatial modes track the drive more closely. The factor $(i)$ reflects the phase lag between the correlation response and the driving.

\section{\label{s3}Excess entropy production}

The excess current is the component of the particle current orthogonal to the instantaneous stationary current. In the weak-drive regime, it determines the excess entropy production rate
\begin{equation}
\dot\Sigma_{\ex}
=
\frac{1}{\chi_0}
\int_0^1 \dd x\, |j_{\ex}(x,t)|^2.
\label{E33}
\end{equation}
Rather than the total entropy production, we consider the excess entropy production because a quasistatic protocol can generate finite nonequilibrium correlations at vanishing excess cost. A genuinely irreversible quantity controlled by $\dot\Sigma_{\ex}$ is the lag of the correlation, which vanishes in the quasistatic limit. For the harmonic protocol, using Eqs.~(\ref{E23}) and (\ref{E25}), the excess current is obtained by subtracting the spatially uniform part, $-\partial_x \rho$. After averaging over one period, we obtain
\begin{equation}
\overline{\dot\Sigma}_{\ex}
=
\frac{\epsilon^2}{2\chi_0}\mathcal F(s),
\qquad
s=\sqrt{\frac{\omega}{2}},
\label{E34}
\end{equation}
where the form factor
\begin{equation}
\mathcal F(s)
=
\frac{
s^2(\cosh s+\cos s)
-s(\sinh s+\sin s)
}{
2(\cosh s-\cos s)
}
\label{E35}
\end{equation}
arises from the explicit spatial integration of $|U_\omega'(x)-1|^2$. The small-frequency expansion reads [see Appendix \ref{F1}]
\begin{equation}
\mathcal F(s)
=
\frac{s^4}{30}
-\frac{s^8}{18900}
+O(s^{12}),
\label{E36}
\end{equation}
so that
\begin{equation}
\overline{\dot\Sigma}_{\ex}
=
\frac{\epsilon^2\omega^2}{240\chi_0}
+O(\omega^4).
\label{E37}
\end{equation}
In the quasistatic limit, $\omega \to 0$, the driving becomes infinitely slow, allowing the system sufficient time to adjust. Consequently, the excess current vanishes, and hence $\overline{\dot\Sigma}_{\ex} \to 0$. Eq.~(\ref{E37}) shows that the excess entropy production vanishes quadratically with the driving frequency.

\begin{figure*}[t]
    \centering
    \includegraphics[width=0.9\textwidth]{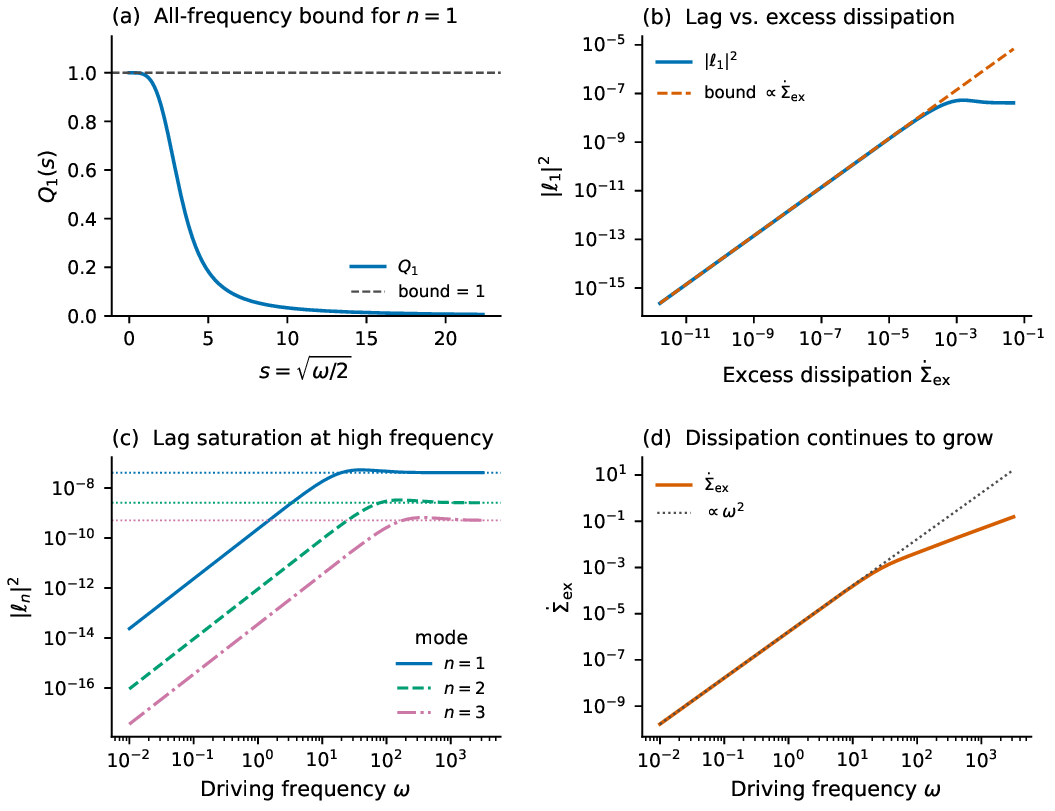}
    \caption{Thermodynamic control of the softest correlation mode. (a) The dimensionless ratio $Q_1(s)$ remains $\leq 1$ for all driving frequencies and approaches unity in the quasistatic limit $s\to 0$.  
(b) Squared lag of the $n=1$ mode as a function of excess entropy production; the dashed line denotes the analytic bound. (c) At high driving frequencies, the lag of each mode saturates to its instantaneous NESS value, while only the $n=1$ mode satisfies the all-frequency bound.  
(d) Excess dissipation continues to increase with frequency, whereas the lag saturates, causing the bound to become progressively looser.}
    \label{fig2}
\end{figure*}

\subsection{Correlation speed limit}
In the quasistatic limit, both the correlation lag and the excess entropy production vanish as $\omega^2$, allowing for a direct comparison between the two quantities. For a given amount of excess entropy production, how much correlation lag can be generated? Remarkably, the slowest mode, $n=1$, obeys a universal bound relating these two quantities. Thus, the correlation lag can never become arbitrarily large compared with the excess dissipation. The slowest spatial mode therefore provides a universal upper bound. For $n=1$ and $\lambda_1=\pi^2$, Eq.~(\ref{E31}) gives
\begin{equation}
|\ell_1(\omega)|^2
=
\frac{4 g_0^2 \epsilon^2 \omega^2 (\omega^2+36 \pi^4)}
{\pi^4 (\omega^2 + 4 \pi^4)(\omega^2 + 16 \pi^4)}.
\label{E38}
\end{equation}
This quantity measures the magnitude of the lag of the slowest correlation mode. Comparing it with Eq.~(\ref{E37}), we define, for the longest-wavelength mode $n=1$,
\begin{equation}
Q_1(\omega)
=
\frac{\pi^8|\ell_1(\omega)|^2}
{540g_0^2\chi_0\overline{\dot\Sigma}_{\ex}(\omega)}.
\label{E39}
\end{equation}

The correlation cannot change rapidly without incurring a thermodynamic cost. A large finite-time correlation lag requires sufficient excess entropy production. If the excess dissipation is constrained to be very small, the correlation cannot deviate significantly from its instantaneous stationary value. This provides the basis for the correlation speed limit.

Substituting Eqs.~(\ref{E37}) and (\ref{E39}) gives
\begin{equation}
Q_1(s)
=
\frac{
4\pi^4s^3(s^4+9\pi^4)
}{
135(s^4+\pi^4)(s^4+4\pi^4)\mathcal{F}(s)
}.
\label{E40}
\end{equation}
The analytical positivity proof is given in Appendix \ref{G1}. It establishes
\begin{equation}
0 < Q_1(s) \le 1,
\qquad s>0,
\label{E41}
\end{equation}
with $Q_1(s)\to 1$ as $s\to 0$. Therefore,
\begin{equation}
|\ell_1(\omega)|^2
\le
\frac{
540g_0^2\rho_0(1-\rho_0)
}{
\pi^8
}
\overline{\dot\Sigma}_{\ex}(\omega).
\label{E42}
\end{equation}
The bound is asymptotically saturated under quasistatic driving. At high frequencies, the correlation cannot follow the drive, and $\ell_1$ approaches the full oscillating instantaneous NESS correlation, whereas the excess dissipation continues to increase.

\subsection{Quasistatic saturation}
From Eq.~(\ref{E32}), we obtain
\begin{equation}
|\ell_1|^2
=
\frac{9g_0^2\epsilon^2\omega^2}{4\pi^8}
+O(\omega^4),
\label{43}
\end{equation}
while Eq.~(\ref{E37}) gives the corresponding low-frequency behavior of the excess entropy production. Therefore,
\begin{equation}
\frac{|\ell_1|^2}{\overline{\dot{\Sigma}}_{\rm ex}}
=
\frac{540g_0^2\chi_0}{\pi^8}
+O(\omega^2)
\label{44}
\end{equation}
Thus, the $n=1$ correlation mode asymptotically saturates the proposed bound in the quasistatic limit, as shown in Fig.~\ref{fig2}(c).

Fig.~\ref{fig2} summarizes the main results. Panel (a) shows that $Q_1$ never exceeds unity and approaches $1$ in the quasistatic limit. Panel (b) presents the same result in the plane of lag versus excess dissipation: the $n=1$ mode lies on or below the bound for all driving frequencies. Panels (c) and (d) illustrate why this bound is special to the softest mode. At high frequencies, all correlation modes saturate to their instantaneous NESS amplitudes, yet only the $n=1$ mode remains controlled by $\dot{\Sigma}_{\mathrm{ex}}$ over the entire frequency range. Higher modes can exceed any analogous bound once $\omega$ becomes comparable to their relaxation rates. Meanwhile, the excess dissipation continues to increase with frequency, causing the inequality to become progressively looser away from the quasistatic regime.
\subsection{Higher modes and High frequency limit}
For a general spatial mode $n$, we define
\begin{equation}
Q_n(\omega)
=
\frac{(n\pi)^8|\ell_n|^2}
{540g_0^2\chi_0\overline{\dot{\Sigma}}_{\rm ex}}.
\label{E45}
\end{equation}
The frequency dependence of the correlation response enters through the mode relaxation rate $\lambda_n=(n\pi)^2$. At the same time, the full diffusive spectrum of the driven density field contributes to the excess entropy production. Consequently, the cancellation that occurs for the softest mode, $n=1$, does not persist for higher modes. This can be demonstrated directly by evaluating $Q_n(\omega)$ at frequencies comparable to the relaxation rates of the higher modes. The resulting behavior shows that the all-frequency bound is a property of the longest-wavelength mode rather than a mode-independent inequality.
\begin{figure}[t]
    \centering
    \includegraphics[width=1.0\columnwidth]{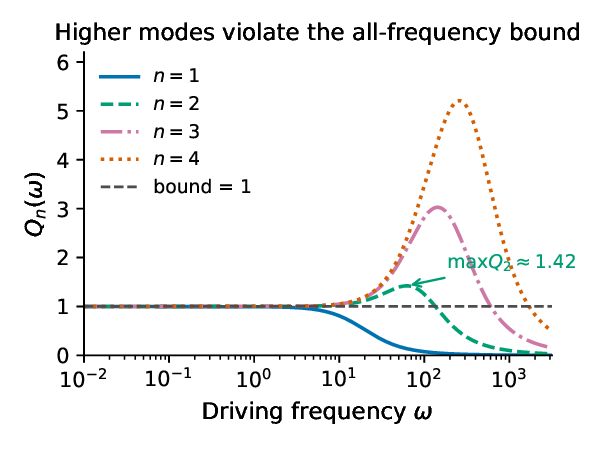}
    \caption{Mode-resolved ratio $Q_n(\omega)$. The horizontal dashed line marks the value 1. While $Q_1(\omega)\le 1$ for all frequencies, the curves for $n=2,3,4$ rise above the bound once $\omega$ becomes comparable to the corresponding mode relaxation rates, confirming that the all-frequency inequality is special to the softest hydrodynamic mode. The peak of the $n=2$ curve reaches $\max Q_2\approx 1.42$.}
    \label{fig3}
\end{figure}

In the high-frequency regime $\omega\gg\lambda_n$, the correlation-mode response decays as $\delta a_n(\omega)=O(\omega^{-2})$, while the corresponding phase lag approaches a finite limiting value, $\ell_n(\omega)=-2g_0\epsilon/\lambda_n+O(\omega^{-1})$. Thus, the correlation field becomes unable to follow the rapidly oscillating instantaneous NESS. In contrast, the excess dissipation increases with frequency because the diffusive density field must respond to increasingly rapid boundary modulation. Consequently, the ratio of the correlation lag to the excess dissipation decreases at high frequency, rendering the corresponding speed bound progressively less restrictive[see Fig.~\ref{fig2}(d)].

Fig.~\ref{fig3} demonstrates that the all-frequency bound is exclusive to the longest-wavelength mode. The dimensionless ratio $Q_n(\omega)$ is constructed such that $Q_1(\omega)\leq 1$ for all frequencies, with equality approached only in the quasistatic limit $\omega\to0$. In contrast, for every higher mode, the corresponding ratio exceeds unity over an intermediate-frequency range. For $n=2$, the ratio reaches a maximum of approximately $1.42$, while the higher modes $n=3$ and $n=4$ exhibit progressively larger overshoots. These maxima occur when the driving frequency becomes comparable to the corresponding mode relaxation rates $\lambda_n=(n\pi)^2$. At both low and high frequencies, the ratios return to values below or equal to unity. The intermediate-frequency overshoot therefore demonstrates that no analogous all-frequency bound controlled solely by the excess dissipation holds for modes with $n\geq2$.

\section{\label{s4}Discussion and conclusion}
The results establish a direct dynamical connection between the nonlocal large-deviation structure of a nonequilibrium steady state and its finite-rate response. For the boundary-driven SSEP the DLS functional determines the stationary long-range correlation kernel. When this kernel is followed under a weak time-dependent boundary protocol, each spatial mode acquires a frequency-dependent lag relative to its instantaneous stationary value. We have shown that the lag of the longest-wavelength mode is constrained by the excess entropy production associated with the driving, while higher modes systematically exceed the corresponding bound.

The distinction between the correlation itself and its dynamical lag is essential. The stationary correlation is a property of the instantaneous nonequilibrium state and can remain finite even when the excess dissipation vanishes. It therefore cannot, by itself, obey a nontrivial bound controlled solely by excess dissipation. The lag, by contrast, vanishes in the quasistatic limit and measures the failure of the evolving state to remain on the instantaneous stationary manifold. It is consequently the appropriate quantity for relating finite-rate response to irreversible cost.

Mode structure further sharpens the result. The \(n=1\) mode is the slowest diffusive mode; its lag satisfies the bound at all frequencies and approaches saturation under quasistatic driving. Higher modes relax on shorter time scales and introduce additional frequency scales into both the response and the dissipation. The precise cancellation that protects the \(n=1\) bound therefore fails for \(n\ge 2\), and the associated dimensionless ratios exceed unity once the driving frequency becomes comparable to the corresponding relaxation rates. The speed limit obtained here is thus a property of the softest correlation mode rather than a mode-independent inequality.

At high frequency every correlation amplitude saturates while excess dissipation continues to grow, so the bound becomes progressively looser. The intermediate-frequency overshoot observed for higher modes is therefore not a generic breakdown of thermodynamic control; it reflects the competition between the external drive and the intrinsic relaxation scale of each mode.

The analysis remains restricted to weak harmonic driving and constant transport coefficients. Whether an analogous bound survives for density-dependent diffusivity and mobility, or for stronger or non-harmonic protocols, is left open. The SSEP nevertheless supplies an analytically controlled setting in which the relation between large-deviation geometry, correlation lag, and excess dissipation can be established explicitly and its mode selectivity resolved.

The nonlocal curvature of the DLS functional has a measurable dynamical consequence: under finite-rate driving its correlation modes acquire frequency-dependent lags whose thermodynamic control depends on their spatial relaxation scale. For the boundary-driven SSEP the longest-wavelength mode obeys an all-frequency bound by excess dissipation, with asymptotic saturation in the quasistatic limit, whereas higher modes can violate the corresponding bound at intermediate frequencies. The result provides a concrete link between large-deviation geometry, finite-frequency correlation dynamics, and excess irreversibility, and shows that such thermodynamic constraints can be intrinsically mode selective.

\appendix

\section{\label{A1}DLS first variation}

For completeness, we define
\begin{equation}
\Lag(\rho,F,F')
=
\rho\ln\frac{\rho}{F}
+(1-\rho)\ln\frac{1-\rho}{1-F}
+\ln\frac{F'}{g}.
\label{A11}
\end{equation}
Because $F$ is determined by stationarity of the variational problem, the envelope theorem gives
\begin{align}
\frac{\delta\Ffun}{\delta\rho}
&=
\frac{\partial\Lag}{\partial\rho}
\label{A12}\\
&=
\ln\frac{\rho}{F}
-\ln\frac{1-\rho}{1-F}
\label{A13}\\
&=
\ln\frac{\rho(1-F)}{F(1-\rho)}.
\label{A14}
\end{align}
At the typical profile, $F=\rho=\bar\rho$, we therefore obtain
\begin{equation}
\left.
\frac{\delta\Ffun}{\delta\rho}
\right|_{\bar\rho}
=0.
\label{A15}
\end{equation}

\section{\label{B1}Linearized auxiliary-field equation}
Let
\begin{equation}
\rho=\bar\rho+\eta,\qquad
F=\bar\rho+u,
\qquad
\bar\rho'=g.
\label{B11}
\end{equation}
Expanding
\begin{equation}
\rho
=
F+F(1-F)\frac{F''}{(F')^2}
\label{B12}
\end{equation}
to first order gives
\begin{equation}
\eta
=
u+
\frac{\bar\rho(1-\bar\rho)}{g^2}u''.
\label{B13}
\end{equation}
Hence, $u$ satisfies
\begin{equation}
\left[
1+\frac{\bar\rho(1-\bar\rho)}{g^2}\partial_x^2
\right]u
=
\eta,
\qquad
u(0)=u(1)=0.
\label{B14}
\end{equation}

The first variation of the functional derivative is
\begin{equation}
\delta\left(
\frac{\delta\Ffun}{\delta\rho}
\right)
=
\frac{\eta-u}{\bar\rho(1-\bar\rho)}.
\label{B15}
\end{equation}
Therefore,
\begin{equation}
\KD
=
\chi^{-1}(I-\mathcal G_F),
\label{B16}
\end{equation}
where $\mathcal G_F$ denotes the Green operator associated with the linearized auxiliary-field equation.

For the SSEP, the inverse operator takes the particularly simple form
\begin{equation}
\KD^{-1}
=
\chi\,\delta
-
g^2(-\partial_x^2)^{-1}_{D}.
\label{B17}
\end{equation}
Since
\begin{equation}
(-\partial_x^2)^{-1}_{D}(x,y)
=
\GD(x,y),
\label{B18}
\end{equation}
Eq.~\eqref{E14} follows.

\section{\label{C1}Spectral representation}

The Dirichlet eigenfunctions satisfy
\begin{equation}
-\partial_x^2\phi_n
=
\lambda_n\phi_n,
\qquad
\lambda_n=(n\pi)^2,
\label{C11}
\end{equation}
and the corresponding Dirichlet Green function admits the spectral representation
\begin{equation}
\GD(x,y)
=
\sum_{n=1}^{\infty}
\frac{\phi_n(x)\phi_n(y)}{\lambda_n}.
\label{C12}
\end{equation}
Substituting this representation into Eq.~\eqref{E14} immediately yields the steady-state mode amplitudes
\begin{equation}
a_n^{\ssstate}
=
\frac{g^2}{\lambda_n}.
\label{C13}
\end{equation}

\section{\label{D1}Dynamical covariance equation}

Write the fluctuating hydrodynamic equation as
\begin{equation}
\partial_t\rho
=
\partial_x^2\rho
-
\partial_x\left[
\sqrt{\frac{2\chi(\rho)}{N}}\,\xi(x,t)
\right],
\label{D11}
\end{equation}
with the noise correlation
\begin{equation}
\langle\xi(x,t)\xi(y,t')\rangle
=
\delta(x-y)\delta(t-t').
\label{D12}
\end{equation}
Linearizing about a deterministic profile gives an Ornstein--Uhlenbeck field. Its covariance therefore obeys
\begin{equation}
\partial_t C
=
(\partial_x^2+\partial_y^2)C
+
2\partial_x\partial_y
\bigl[\chi(\rho)\delta(x-y)\bigr].
\label{D13}
\end{equation}

We decompose the covariance into its local equilibrium contribution and its long-range part,
\begin{equation}
C(x,y,t)
=
\chi(\rho(x,t))\delta(x-y)
+
B(x,y,t).
\label{D14}
\end{equation}
The long-range part $B$ then satisfies
\begin{equation}
\partial_t B
=
(\partial_x^2+\partial_y^2)B
+
2\partial_x\partial_y
\bigl[\chi(\rho)\delta(x-y)\bigr]
-
\partial_t
\bigl[\chi(\rho)\delta(x-y)\bigr].
\label{D15}
\end{equation}
For the weak-drive harmonic calculation, terms beyond the required orders in $g_0$ and $\epsilon$ are discarded after the deterministic solution is inserted.

\section{\label{E1}Harmonic deterministic profile}

For the harmonic boundary protocol
\begin{equation}
g(t)=g_0+\epsilon\cos(\omega t),
\label{E111}
\end{equation}
the complex amplitude $U_\omega(x)$ satisfies
\begin{equation}
U_\omega''
=
i\omega U_\omega,
\qquad
U_\omega(0)=-\frac{1}{2},
\qquad
U_\omega(1)=\frac{1}{2}.
\label{E112}
\end{equation}
Thus,
\begin{equation}
U_\omega(x)
=
\frac{\sinh[k(x-\tfrac{1}{2})]}
{2\sinh(k/2)},
\qquad
k^2=i\omega.
\label{E113}
\end{equation}
The corresponding spatial derivative is
\begin{equation}
U_\omega'(x)
=
\frac{k\cosh[k(x-\tfrac{1}{2})]}
{2\sinh(k/2)}.
\label{E114}
\end{equation}

The mode projection required for the covariance response can be evaluated using the identity
\begin{equation}
\int_0^1 dx\,
U_\omega'(x)\phi_n^2(x)
=
\frac{4\lambda_n}{4\lambda_n+i\omega}.
\label{E115}
\end{equation}
Inserting this result into the projected covariance equation gives
\begin{equation}
\delta a_n(\omega)
=
\frac{16g_0\epsilon\lambda_n}
{(2\lambda_n+i\omega)(4\lambda_n+i\omega)}.
\label{E116}
\end{equation}
This is the finite-frequency correlation response quoted in Eq.~(\ref{E27}) of the main text.

In the quasistatic limit, the corresponding instantaneous stationary response is
\begin{equation}
\delta a_n^{\rm ss}
=
\frac{2g_0\epsilon}{\lambda_n}.
\label{E117}
\end{equation}
Subtracting Eq.~\eqref{E117} from Eq.~\eqref{E116} gives the correlation lag defined in Eq.~(\ref{E29}) of the main text.

\section{\label{F1}Low-frequency expansion}
Using the Taylor expansions
\begin{align}
\cosh s+\cos s
&=
2+\frac{s^4}{12}
+\frac{s^8}{20160}
+\cdots,
\label{F11}\\
\sinh s+\sin s
&=
2s+\frac{s^5}{60}
+\frac{s^9}{181440}
+\cdots,
\label{F12}\\
\cosh s-\cos s
&=
s^2+\frac{s^6}{360}
+\frac{s^{10}}{1814400}
+\cdots,
\label{F13}
\end{align}
and substituting them into the definition of the form factor $\mathcal F(s)$ in Eq.~(\ref{E35}) of the main text, one obtains
\begin{equation}
\mathcal F(s)
=
\frac{s^4}{30}
-\frac{s^8}{18900}
+O(s^{12}).
\label{F14}
\end{equation}
Since
\begin{equation}
s^2=\frac{\omega}{2},
\label{F15}
\end{equation}
the excess entropy production has the low-frequency expansion
\begin{equation}
\overline{\dot\Sigma}_{\ex}
=
\frac{\epsilon^2\omega^2}{240\chi_0}
+O(\omega^4).
\label{F16}
\end{equation}
This reproduces Eq.~(\ref{E37}) of the main text.

\section{\label{G1}Proof of the $n=1$ bound}

Substituting Eqs.~(\ref{E37}) and (\ref{E39}) into the definition of $Q_1$ and simplifying gives Eq.~(\ref{E40}) of the main text. The desired inequality is
\begin{equation}
Q_1(s)\leq 1.
\label{G11}
\end{equation}

Define
\begin{align}
A(s) &= s(\cosh s+\cos s) -(\sinh s+\sin s),
\label{G12}\\
B(s)&= \cosh s-\cos s.
\label{G13}
\end{align}

For $s>0$, both $A(s)$ and $B(s)$ are positive. Indeed,
\begin{equation}
A'(s) = s(\sinh s-\sin s)>0,
\qquad
A(0)=0,
\label{G14}
\end{equation}
because $\sinh s>\sin s$ for $s>0$, while
\begin{equation}
B'(s) = \sinh s+\sin s>0.
\label{G15}
\end{equation}

The inequality in Eq.~\eqref{G11} is equivalent to
\begin{equation}
H(s)\geq 0,
\label{G16}
\end{equation}
where
\begin{align}
H(s)
={}&
135(s^4+\pi^4)(s^4+4\pi^4)A(s)
\nonumber\\
&-
4\pi^4s^3(s^4+9\pi^4)B(s).
\label{G17}
\end{align}

Both $A(s)$ and $B(s)$ vanish at $s=0$ to the required orders. Expanding in powers of $s$ gives
\begin{equation}
H(s)
=
\sum_{m\geq0}h_m s^{4m+9},
\label{G18}
\end{equation}
with
\begin{equation}
h_m>0.
\label{G19}
\end{equation}

The coefficients follow by inserting the absolutely convergent Taylor series
\begin{align}
A(s)
&=
\sum_{m=1}^{\infty}
\frac{2s^{4m+1}}{(4m-1)!},
\label{G20}\\
B(s)
&=
2\sum_{m=0}^{\infty}
\frac{s^{4m+2}}{(4m+2)!}.
\label{G21}
\end{align}

After collecting equal powers, the coefficient can be written as 
\begin{equation}
h_m
=
\frac{P_m(\pi^4)}
{(4m+9)!},
\label{G22}
\end{equation}
where $P_m(z)$ is a polynomial with positive coefficients for $z=\pi^4$. Hence,
\begin{equation}
h_m>0,
\label{G23}
\end{equation}
and therefore
\begin{equation}
H(s)>0
\qquad
\text{for every }s>0.
\label{G24}
\end{equation}

Consequently,
\begin{equation}
Q_1(s)<1,
\qquad
s>0,
\label{G25}
\end{equation}
and continuity gives
\begin{equation}
Q_1(0)=1.
\label{G26}
\end{equation}
This establishes the bound stated in Eq.~(\ref{E41}) and, consequently, the correlation speed limit in Eq.~(\ref{E42}) of the main text.

\bibliography{bibliography}

@article{Spohn1983,
  author  = {Spohn, H.},
  title   = {Long range correlations for stochastic lattice gases in a non-equilibrium steady state},
  journal = {Journal of Physics A: Mathematical and General},
  volume  = {16},
  number  = {18},
  pages   = {4275--4291},
  year    = {1983},
  doi     = {10.1088/0305-4470/16/18/029}
}

@article{Bertini2015,
  author  = {Bertini, L. and De Sole, A. and Gabrielli, D. and Jona-Lasinio, G. and Landim, C.},
  title   = {Macroscopic fluctuation theory},
  journal = {Reviews of Modern Physics},
  volume  = {87},
  number  = {2},
  pages   = {593--636},
  year    = {2015},
  doi     = {10.1103/RevModPhys.87.593}
}

@article{Derrida2001,
  author  = {Derrida, B. and Lebowitz, J. L. and Speer, E. R.},
  title   = {Free energy functional for nonequilibrium systems: An exactly solvable case},
  journal = {Physical Review Letters},
  volume  = {87},
  number  = {15},
  pages   = {150601},
  year    = {2001},
  doi     = {10.1103/PhysRevLett.87.150601}
}

@article{Derrida2002,
  author  = {Derrida, B. and Lebowitz, J. L. and Speer, E. R.},
  title   = {Large deviation of the density profile in the steady state of the open symmetric simple exclusion process},
  journal = {Journal of Statistical Physics},
  volume  = {107},
  number  = {3--4},
  pages   = {599--634},
  year    = {2002},
  doi     = {10.1023/A:1014555927320}
}

@article{Bodineau2024,
  author  = {Bodineau, T. and Dagallier, B.},
  title   = {Large deviations for out of equilibrium correlations in the symmetric simple exclusion process},
  journal = {Electronic Journal of Probability},
  volume  = {29},
  number  = {63},
  pages   = {1--96},
  year    = {2024},
  doi     = {10.1214/24-EJP1121}
}

@article{Hatano2001,
  author  = {Hatano, T. and Sasa, S.-i.},
  title   = {Steady-state thermodynamics of Langevin systems},
  journal = {Physical Review Letters},
  volume  = {86},
  number  = {16},
  pages   = {3463--3466},
  year    = {2001},
  doi     = {10.1103/PhysRevLett.86.3463}
}

@article{Kolchinsky2026,
  author  = {Kolchinsky, Artemy and Dechant, Andreas and Yoshimura, Kohei and Ito, Sosuke},
  title   = {Generalized free energy and excess/housekeeping decomposition in nonequilibrium systems: From large deviations to thermodynamic speed limits},
  journal = {Physical Review Research},
  volume  = {8},
  number  = {2},
  pages   = {023025},
  year    = {2026},
  doi     = {10.1103/r48t-dghl}
}

@article{Dechant2023,
  author  = {Dechant, Andreas and Garnier-Brun, J{\'e}r{\^o}me and Sasa, Shin-ichi},
  title   = {Thermodynamic bounds on correlation times},
  journal = {Physical Review Letters},
  volume  = {131},
  number  = {16},
  pages   = {167101},
  year    = {2023},
  doi     = {10.1103/PhysRevLett.131.167101}
}

@article{Shiraishi2018,
  author  = {Shiraishi, Naoto and Funo, Ken and Saito, Keiji},
  title   = {Speed limit for classical stochastic processes},
  journal = {Physical Review Letters},
  volume  = {121},
  number  = {7},
  pages   = {070601},
  year    = {2018},
  doi     = {10.1103/PhysRevLett.121.070601}
}

@article{Bertini2002,
  author  = {Bertini, L. and De Sole, A. and Gabrielli, D. and Jona-Lasinio, G. and Landim, C.},
  title   = {Macroscopic fluctuation theory for stationary non-equilibrium states},
  journal = {Journal of Statistical Physics},
  volume  = {107},
  number  = {3--4},
  pages   = {635--675},
  year    = {2002},
  doi     = {10.1023/A:1014525911391}
}

@article{Maes2014,
  author  = {Maes, Christian and Netocny, Karel},
  title   = {A nonequilibrium extension of the Clausius heat theorem},
  journal = {Journal of Statistical Physics},
  volume  = {154},
  number  = {1--2},
  pages   = {188--203},
  year    = {2014},
  doi     = {10.1007/s10955-013-0822-9}
}

@article{Bertini2003,
  author  = {Bertini, L. and De Sole, A. and Gabrielli, D. and Jona-Lasinio, G. and Landim, C.},
  title   = {Large deviations for the boundary driven symmetric simple exclusion process},
  journal = {Mathematical Physics, Analysis and Geometry},
  volume  = {6},
  number  = {3},
  pages   = {231--267},
  year    = {2003},
  doi     = {10.1023/A:1024967818899}
}
\end{document}